\documentclass[12pt]{article}
\usepackage{amsmath}
\usepackage{fullpage}
\usepackage{graphicx}
\newcommand{\be}{\begin{equation}}
\newcommand{\ee}{\end{equation}}

\font\mybb=msbm10 at 11pt

\def\bb#1{\hbox{\mybb#1}}

\newcommand{\news}{\setcounter{equation}{0}\quad}
\def\ben{\begin{equation}}
\def\een{\end{equation}}
\def\bea{\begin{eqnarray}}
\def\eea{\end{eqnarray}}
\begin{document}
\title{
\begin{flushright}\ \vskip -2cm {\normalsize{\em DCPT-09/49}}\end{flushright}\vskip 1cm
\large {\bf STABILITY AND THE EQUATION OF STATE \\ FOR KINKY VORTONS}}
\author{
Richard A. Battye$^{1}$
and Paul M. Sutcliffe$^{2}$\\[10pt]
\\{\normalsize $^{1}$
{\sl Jodrell Bank Centre for Astrophysics,} }
\\{\normalsize {\sl University of Manchester,
 Manchester M13 9PL, U.K.}}
\\{\normalsize {\sl Email : Richard.Battye@manchester.ac.uk}}\\
\\{\normalsize $^{2}$
{\sl \normalsize Department of Mathematical Sciences,
Durham University, Durham DH1 3LE, U.K.}}\\
{\normalsize {\sl Email: p.m.sutcliffe@durham.ac.uk}}\\[10pt]}
\date{August 2009}
\maketitle
\begin{abstract}
Vortons are closed loops of superconducting strings carrying
current and charge. A formalism has been developed to study vortons
in terms of an elastic string approximation, but its implementation 
requires knowledge of the unknown equation of state, relating the string
tension to the energy per unit length. 
Recently, a planar analogue of the vorton, known as a kinky vorton, has
been introduced. In this paper we derive an exact formula
for the equation of state of a kinky vorton and use it to calculate
the properties of the associated elastic string, such as the
transverse and longitudinal propagation speeds. In particular, the
elastic string approximation predicts a complicated and highly
non-trivial pattern of intervals of instability, which we are able
to confirm using full field simulations. The implications of the
results for vortons are also discussed.
   
\end{abstract}

\newpage

\section{Introduction}\news
Superconducting cosmic strings were introduced by Witten \cite{Wi},
who realized that if the field of a cosmic string is coupled to
another complex scalar field then a non-dissipative current can
flow along the string. A vorton \cite{DS2} is a closed loop of
superconducting cosmic string that carries both current and charge,
which may provide a force to balance the string tension and prevent 
its collapse. As vortons have a number of possible cosmological consequences
\cite{BCDT},  it is of considerable interest to determine their
stationary and dynamical properties.  

Recently, results have been presented \cite{BS2} on the first numerical 
construction of vortons in the simplest theory, namely, the global
version of Witten's $U(1)\times U(1)$ theory. These results demonstrate
that a range of stationary circular vortons exist, that are stable
to axially symmetric perturbations. However, unstable modes associated
with non-axial perturbations were found to exist. 

A formalism has been developed \cite{Ca,CM} to study vortons and
their stability
in terms of an elastic string approximation, but its implementation 
requires knowledge of the unknown equation of state, relating the string
tension to the energy per unit length. Attempts have been made \cite{Pe} to
compute the equation of state numerically from field theory
simulations using particular values of the field theory parameters, but
obviously this is far from ideal.

The equation of state can be derived
from the action density on the string worldsheet, but one needs to know
how this depends on the quantity $\chi=\omega^2-k^2,$ 
where $\omega$ and $k$ are the frequency and twist rate of the phase 
of the condensate field. Witten originally
suggested \cite{Wi} that the usual Nambu Goto action density, which is simply
a constant, could be replaced by a form linear in $\chi.$ 
However, this linear form produces results which are qualitatively incorrect,
even for small $\chi,$ since the propagation speed of longitudinal 
perturbations depends on the second derivative of the action density
with respect to $\chi.$ This prompted suggestions \cite{CP,Ca2,HC} 
for alternative nonlinear forms for the assumed equation of state.

Recently, a planar analogue of the vorton, known as a kinky vorton, has
been introduced and shown to possess many of the features expected
of a vorton \cite{BS}. A significant advantage of the kinky vorton is
that several approximations required in the study of vortons can be 
replaced by exact results. In this paper we exploit this advantage 
to derive an exact formula for the kinky vorton equation of state, and 
make a comparison with the previously proposed approximations to
the vorton equation of state. The exact equation of state is then
used to calculate properties, such as the transverse and longitudinal 
propagation speeds, of the elastic string description of the kinky vorton.
In particular, the elastic string model predicts a complicated and highly
non-trivial pattern of intervals of instability, which we are able
to confirm using full field simulations. These analytic results provide an
understanding of similar instabilities found for vortons using numerical 
simulations of the full field theory dynamics \cite{BS2}.

\section{The action density on the string worldsheet}\news
The kinky vorton Lagrangian density in (2+1)-dimensions is given by
\cite{BS}
\be
{\cal L}=\partial_\mu\phi \partial^\mu \phi
+\partial_\mu \sigma \partial^\mu \bar\sigma
-\frac{\lambda_\phi}{4}(\phi^2-\eta_\phi^2)^2
-\frac{\lambda_\sigma}{4}(|\sigma|^2-\eta_\sigma^2)^2
-\beta\phi^2 |\sigma|^2+\frac{\lambda_\sigma}{4}\eta_\sigma^4
\label{lagden}
\ee
where $\phi$ and $\sigma$ are real and complex scalar fields respectively,
with $\eta_\phi,\eta_\sigma,\lambda_\phi,\lambda_\sigma,\beta$ all 
real positive constants.

This theory can be obtained from the global version of Witten's
$U(1)\times U(1)$ theory \cite{Wi} by a trivial dimensional reduction
from (3+1)-dimensions to (2+1)-dimensions, followed by a restriction
that one of the complex scalar fields is real.

The theory has a global $\bb{Z}_2\times U(1)$ symmetry and the parameters
of the model can be arranged so that in the vacuum the $\bb{Z}_2$ symmetry is
broken, $\phi=\pm\eta_\phi\ne 0,$ while the $U(1)$ symmetry remains unbroken,
$|\sigma|=0.$
For this symmetry breaking pattern there exist kink strings constructed
from the $\phi$ field. If  the infinite kink string lies along the $y$-axis,
then it is given by the solution
\be
\phi=\eta_\phi\tanh\bigg(\frac{\eta_\phi\sqrt{\lambda_\phi}x}{2}\bigg), \quad
\sigma =0.
\label{barekink}
\ee
The situation of interest is when a condensate of the $\sigma$ field
carrying current and charge forms in the core of the kink string.
For the infinite string given above, such a condensate field takes the form
\be
\sigma=e^{i(\omega t+ky)}|\sigma|,
\label{condensate}
\ee
where $|\sigma|$ is a function of $x$ only with
$|\sigma|\rightarrow 0$ as $|x|\rightarrow\infty.$
The constant $k$ describes the rate of twisting of the condensate
along the string, though in the literature this is referred to as winding,
rather than twisting, so we will stick to this common convention.

A non-zero value of $\omega$ induces a charge $Q$ associated with the
global $U(1)$ symmetry, and the winding $k$ generates a current along
the string.
It is easy to see that charge and current will have opposite effects on
the string, so it is useful to introduce the combination
\be
\chi\equiv\omega^2-k^2.
\label{chi}
\ee
In the literature solutions with $\chi=0$ are termed chiral, whereas
solutions with $\chi>0$ are referred to as electric and those with
$\chi<0$ are called magnetic \cite{CP,LS}.

To obtain exact solutions for the whole range of $\chi$ we set 
\cite{BS}
\be
2\beta=\lambda_\phi=\lambda_\sigma\equiv\lambda. \label{paramx} \ee
With this choice there are three remaining parameters, which 
may be chosen to be $\eta_\phi$ and the two combinations
\be
m_\phi^2\equiv\frac{\lambda}{2}\eta_\phi^2, \quad 
\alpha\equiv\bigg(\frac{\eta_\sigma}{\eta_\phi}\bigg)^2.
\label{mass}
\ee
In fact, the parameters $m_\phi$ and $\eta_\phi$ merely set energy
and length units and can therefore be set to unity without loss of generality.
To restore these parameters in the following the scalings
$L\mapsto m_\phi \eta_\phi^2 L$ and $\chi\mapsto \chi/m_\phi^2$ 
need to be applied.

The only significant parameter that remains is $\alpha$, which must
lie in the interval $\alpha\in(\frac{1}{2},1)$ if chiral solutions 
are to exist.
The range of $\chi$ is 
\be
\chi_-\equiv \frac{1}{2}-\alpha \le \chi \le 1-\alpha \equiv \chi_+,
\label{range}
\ee
and the associated exact solutions are
\be
\phi={\rm tanh}(x\sqrt{1-\alpha-\chi}), \quad \quad
\sigma=e^{i(\omega t+ky)}
\sqrt{2\alpha-1+2\chi}\,{\rm sech}(x\sqrt{1-\alpha-\chi}).
\label{exact}
\ee

Substituting the solution (\ref{exact}) into the Lagrangian
density (\ref{lagden}) and integrating over $x$ yields the result
\be
L=\frac{4}{3\sqrt{1-\alpha-\chi}}
\bigg\{
(2\alpha+1)(\alpha-1)+\chi(2\chi+4\alpha-1)
\bigg\}
\label{lag}
\ee
for the action density on the string worldsheet.

This should be contrasted vith various proposed forms, such as
the linear approximation
\be
L_{\rm lin}=-m^2+\kappa \chi,
\label{lin}
\ee
due to Witten \cite{Wi}, and
the logarithmic approximation \cite{CP,HC}
\be
L_{\rm log}=-m^2-\frac{1}{2}m_*^2\log(1-\delta_*^2 \chi).
\label{log}
\ee

The parameters $m^2$ and $\kappa$ in the linear approximation
(\ref{lin}) can be determined from the exact expression (\ref{lag})
by matching the function and the derivative at $\chi=0.$
This gives
\be
m^2=\frac{4}{3}(2\alpha+1)\sqrt{1-\alpha}, \quad\quad
\kappa=\frac{4\alpha-2}{\sqrt{1-\alpha}}.
\ee
One approach to fixing the additional parameters 
$m_*^2$ and $\delta_*^2$ in the logarithmic approximation is by
matching the derivative at $\chi=0,$ and requiring the singularity
in (\ref{log}) to occur at the extreme electric limit $\chi=\chi_+,$
which gives
\be
m_*^2=4(2\alpha-1)\sqrt{1-\alpha}, \quad\quad \delta_*^2=1/(1-\alpha).
\ee 
\begin{figure}
\begin{center}
\includegraphics[width=12cm]{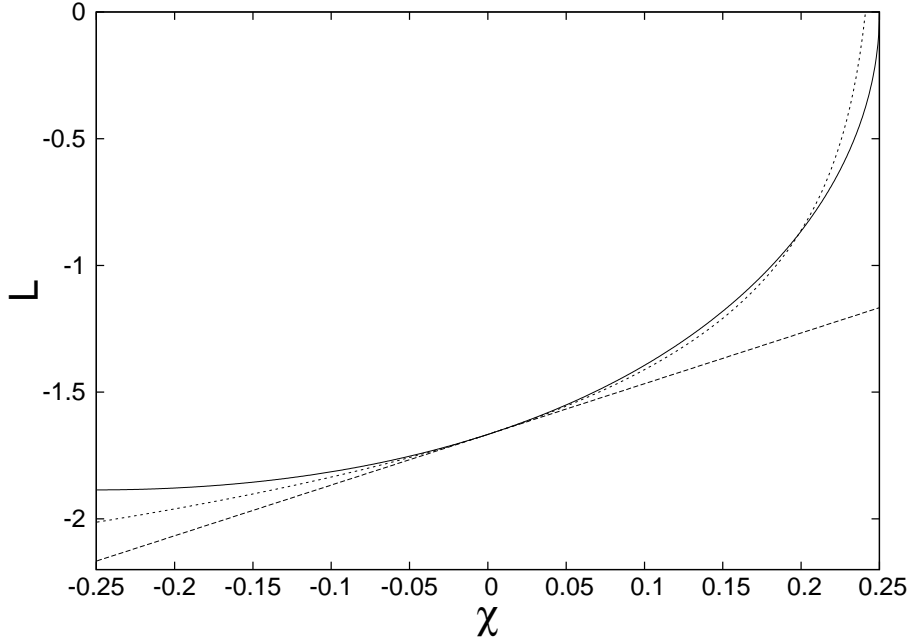}
\caption{The exact action density on the worldsheet $L$ (solid curve)
as a function of $\chi,$ 
together with the linear approximation $L_{\rm lin}$ (dashed curve) 
and the logarithmic approximation $L_{\rm log}$ (dotted curve).
 This is in the theory with $\alpha=3/4.$
}
\label{fig-lag}
\end{center}
\end{figure}

In Figure~\ref{fig-lag} the exact result (\ref{lag}) is compared with the
linear (\ref{lin}) and logarithmic (\ref{log})
approximations for the parameter value
$\alpha=3/4,$ with other parameter values producing similar results.
This shows that the linear approximation is poor, but the logarithmic
approximation is reasonably good, and therefore should produce 
results which are, at least, qualitatively correct.

\section{The equation of state}\news
From the Lagrangian (\ref{lag}) the string tension $T$ and energy
per unit length $U$ can be calculated as follows \cite{Ca}.

The energy-momentum tensor associated with the Lagrangian density
(\ref{lagden}) is
\be
{\cal T}^\mu_\nu=2g^{\mu\alpha}\frac{\partial {\cal L}}{\partial g^{\alpha\nu}}
-\delta^\mu_\nu{\cal L}.
\label{emon}
\ee
Integration over the string cross-section gives the macroscopic tensor
\be
T^{ab}=\int_{-\infty}^\infty {\cal T}^{ab} \, dx,
\label{macro}
\ee
where $a,b\in\{t,y\}.$ 
The energy per unit length $U$ and the string tension $T$ are the
eigenvalues of $T^{ab},$ and are given by $U=T^{tt}$ and $T=-T^{yy},$
in a frame in which $T^{ab}$ is diagonal.   

Applying the above procedure to the Lagrangian density (\ref{lagden})
produces
\be
T^{tt}=2\omega^2\Sigma_2-L, \quad\quad
-T^{yy}=-2k^2\Sigma_2-L,
\label{nondiag}
\ee
where 
\be\Sigma_2=\int_{-\infty}^\infty |\sigma|^2 \, dx
=\frac{4\chi+4\alpha-2}{\sqrt{1-\alpha-\chi}}.
\label{sigma2}
\ee

The frame in which $T^{ab}$ is diagonal is obtained by setting $k=0,$
if $\chi\ge 0,$ or by setting $\omega=0$ if $\chi\le 0.$

In the electric regime, that is $\chi>0,$ setting $k=0$ implies that
$\chi=\omega^2$ and then (\ref{nondiag}) becomes
\be
U=2\chi\Sigma_2-L, \quad\quad\quad T=-L.
\label{elec1}
\ee
Conversely, in the magnetic regime, that is $\chi<0,$ then setting
$\omega=0$ means that $\chi=-k^2$ and hence (\ref{nondiag}) gives
\be
U=-L, \quad\quad\quad  T=2\chi\Sigma_2-L.
\label{mag1}
\ee
Of course, both forms agree in the chiral limit $\chi=0$ where
$U=T=-L.$

Using the explicit expressions (\ref{lag}) and (\ref{sigma2}) for
$L$ and $\Sigma_2$ the above equations become
\be
T=
\begin{cases}
\frac{4}{3}
\bigg(
(2\alpha+1)(1-\alpha)-\chi(2\chi+4\alpha-1)
\bigg)/{\sqrt{1-\alpha-\chi}} & \text{if $\chi\ge 0$} \\
\frac{4}{3}
\bigg(
(2\alpha+1)(1-\alpha)+2\chi(2\chi+\alpha-1)
\bigg)/{\sqrt{1-\alpha-\chi}}
  &  \text{if $\chi<0$}\\
\end{cases}
\label{tension}
\ee
and
\be
U=
\begin{cases}
\frac{4}{3}
\bigg(
(2\alpha+1)(1-\alpha)+2\chi(2\chi+\alpha-1)
\bigg)/{\sqrt{1-\alpha-\chi}}
  &  \text{if $\chi\ge 0$}\\
\frac{4}{3}
\bigg(
(2\alpha+1)(1-\alpha)-\chi(2\chi+4\alpha-1)
\bigg)/{\sqrt{1-\alpha-\chi}} & \text{if $\chi< 0$}. \\
\end{cases}
\label{energy}
\ee
Using these formulae
it is easy to calculate the behaviour 
of the tension and energy in
the three limits $\chi=0$ (chiral), $\chi\rightarrow \chi_+$ (extreme
electric) and  $\chi\rightarrow \chi_-$ (extreme magnetic).  

If $\chi=0$ then $T=U=m^2=\frac{4}{3}(2\alpha+1)\sqrt{1-\alpha}.$
\ As  $\chi\rightarrow \chi_+$ then $T\rightarrow 0$ and $U\rightarrow\infty.$
Finally, as  $\chi\rightarrow \chi_-$ then $T\rightarrow 4\sqrt{2}/3$ and 
$U\rightarrow 4\sqrt{2}/3,$ so again the tension and energy are equal,
and independent of $\alpha.$

\begin{figure}
\begin{center}
\includegraphics[width=12cm]{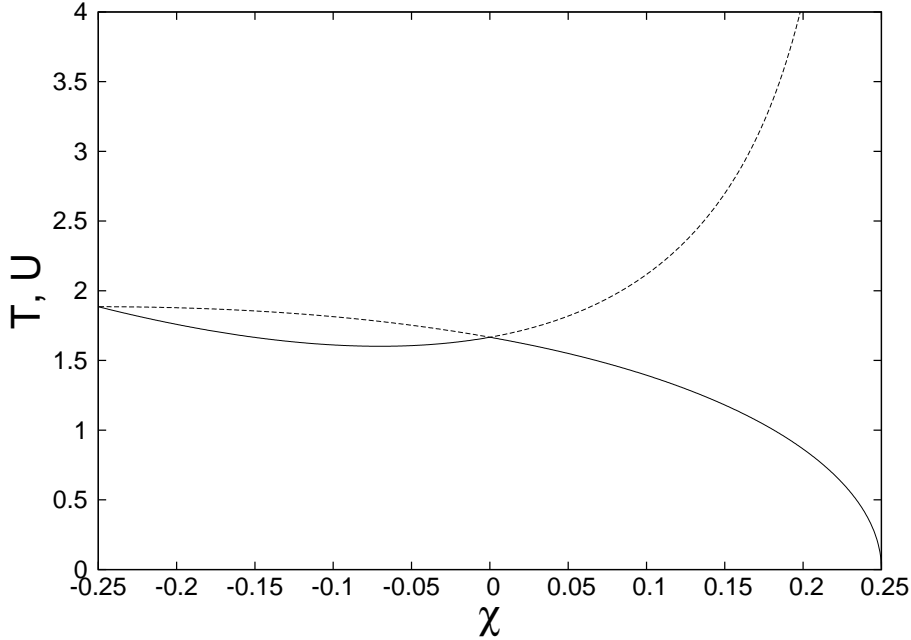}
\caption{The tension $T$ (solid curve) and the energy per unit length
$U$ (dashed curve) as a function of $\chi,$ for the theory with 
$\alpha=3/4.$
}
\label{fig-tu}
\end{center}
\end{figure}

Figure~\ref{fig-tu} displays the tension 
$T$ (solid curve) and the energy per unit length
$U$ (dashed curve) as a function of $\chi,$ for the theory with 
$\alpha=3/4.$ This value of $\alpha$ has been employed in previous
work since the range of $\chi$ is then symmetric around zero,
explicitly $\chi_+=1/4$ and $\chi_-=-1/4,$ allowing a range of electric
and magnetic kinky vortons. Plots for other values of $\alpha$ share the
same qualitative features. All graphs in this paper are plotted for the
theory with $\alpha=3/4,$ but again the qualitative features are independent
of $\alpha.$

Note that the tension is never negative, which agrees with numerical
computations in (3+1)-dimensions, confirming that 
there are no spring states \cite{Pe2}, 
despite earlier claims in the literature (see the discussion and 
references in \cite{Pe2}).

The equation of state is the relation between $T$ and $U,$ and this
is given implicitly by combining the formulae (\ref{tension}) and 
(\ref{energy}). 
\begin{figure}[ht]
\begin{center}
\includegraphics[width=12cm]{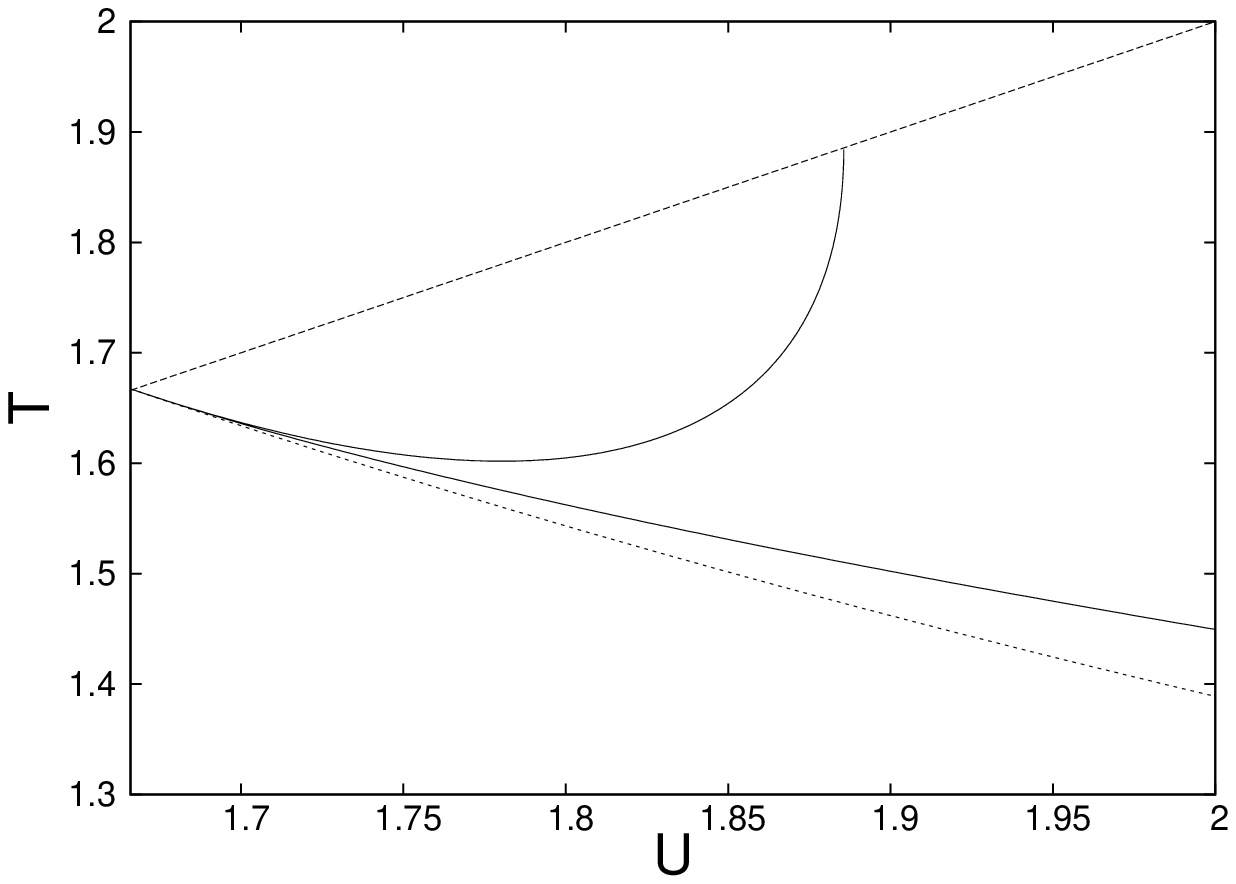}
\caption{The tension $T$ as a function of the energy per unit length 
$U$ (solid curve), for $\alpha=3/4.$ The upper portion of the curve
is the magnetic regime ($\chi<0$) and the lower portion is the
electric regime ($\chi>0$). For comparison the upper dashed line
is the Nambu-Goto equation of state $T=U,$ and the lower dotted curve
is the self-dual equation of state $T=m^4/U.$ 
}
\label{fig-tvu}
\end{center}
\end{figure}
Figure~\ref{fig-tvu} presents the equation of state in graphical form,
that is, $T$ as a function of $U.$ For comparison the upper dashed line
is the Nambu-Goto equation of state $T=U,$ and the lower dotted curve
is the self-dual equation of state \cite{Ca2} $T=m^4/U,$ with
the characteristic property that the equation of state is identical
in the magnetic and electric regimes. The self-dual model appears to
be a reasonable approximation in the electric regime, but as 
discussed in \cite{CP} and demonstrated below, the self-dual
model does not capture the important qualitative features associated
with propagation speeds, even in the electric regime.

Figure~\ref{fig-tvu} has a remarkable similarity to the 
graphical equation of state computed numerically 
for superconducting cosmic strings in (3+1)-dimensions \cite{Pe}.
This suggests that our explicit exact formulae derived in (2+1)-dimensions
also provide a good description of the (3+1)-dimensional system, where
results are only available numerically. Note that the most interesting
dynamics of a string loop takes place in the plane of the loop, therefore
the reduction to (2+1)-dimensions is still expected to capture the most
important degrees of freedom.

String dynamics depends crucially on the
propagation speeds of transverse and longitudinal (sound-like) perturbations.
The equation of state allows the calculation 
of the transverse speed $c_T$ and the longitudinal speed $c_L$ via the 
formulae
\be
c_T^2=\frac{T}{U}, \quad\quad c_L^2=-\frac{dT}{dU}.
\label{speeds}
\ee
Substitution of the tension and energy expressions (\ref{tension}) 
and (\ref{energy}) into the speed formulae (\ref{speeds}) gives
\be
c_T^{2\,{\rm sign}(\chi)}=
\frac{(2\alpha+1)(1-\alpha)-\chi(2\chi+4\alpha-1)}   
{(2\alpha+1)(1-\alpha)+2\chi(2\chi+\alpha-1)},
\label{ct}
\ee  
\be
c_L^{2\,{\rm sign}(\chi)}=
\frac{(2\alpha+2\chi-1)(1-\alpha-\chi)}   
{(2\alpha-1)(1-\alpha)-2\chi(2\chi+3\alpha-3)},
\label{cl}
\ee 
with $c_T=c_L=1$ if $\chi=0.$

The string is unstable if either of the speeds are imaginary, that is,
if either $c_T^2<0$ or $c_L^2<0.$ It is easy to see that $c_T^2>0$ 
for all $\chi\in(\chi_-,\chi_+),$ and $c_T\rightarrow 0$ as
$\chi\rightarrow \chi_+.$ Similarly, it is easy to show that
$c_L^2>0$ for all $\chi\in[0,\chi_+),$ and again $c_L\rightarrow 0$ as
$\chi\rightarrow \chi_+.$ However, $c_L^2<0$ if $\chi<\chi_c,$ where
the critical value for the onset of instability is given by the
vanishing of the denominator in (\ref{cl}), leading to
\be
\chi_c=\frac{3}{4}(1-\alpha)-\frac{1}{4}\sqrt{(1-\alpha)(5-\alpha)}.
\label{critical}
\ee
For the parameter value $\alpha=3/4$ this gives $\chi_c=(3-\sqrt{17})/16
\approx-0.07.$ This is in excellent agreement with numerical field
theory simulations, performed for the value $\alpha=3/4,$ which
estimated the critical value to be $-0.08$ \cite{BS}.

\begin{figure}[ht]
\begin{center}
\includegraphics[width=12cm]{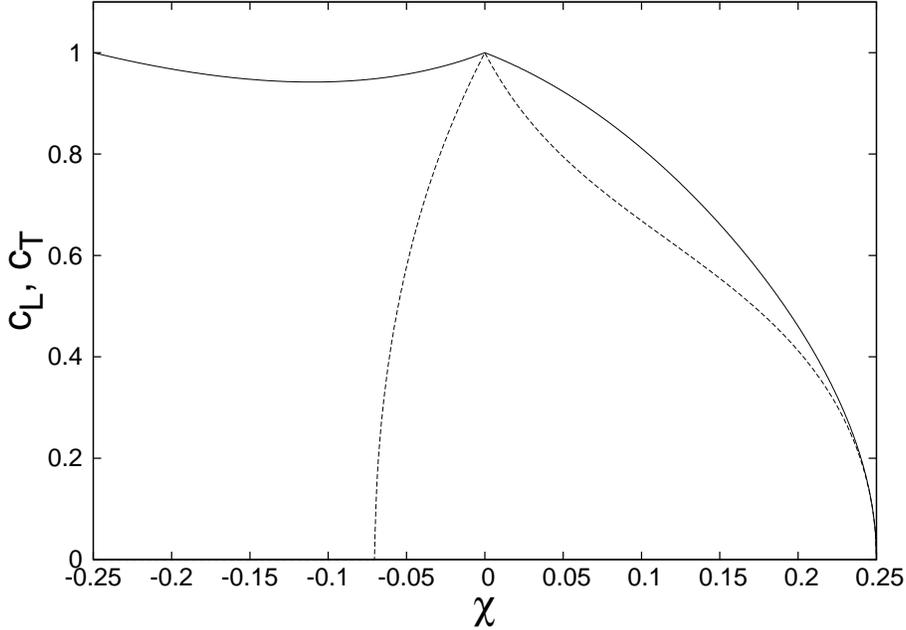}
\caption{
The transverse speed $c_T$ (solid curve) and
the longitudinal speed $c_L$ (dashed curve)
as a function of $\chi,$ for the theory with $\alpha=3/4.$
}
\label{fig-speeds}
\end{center}
\end{figure}

In Figure~\ref{fig-speeds} the speeds $c_T$ (solid curve)
and $c_L$ (dashed curve) are plotted as a function of $\chi,$
for $\alpha=3/4.$ Plots with other values of $\alpha$ are
qualitatively similar. 

In the chiral limit $\chi=0,$ and the extreme electric limit 
$\chi\rightarrow\chi_+,$ the string is transonic, that is, $c_T=c_L.$
For all other values of $\chi$ the string is supersonic, that is,
$c_T>c_L.$ Such supersonic behaviour has been observed in 
numerical computations based on (3+1)-dimensional field theories \cite{Pe}.

The simple linear approximation to the action (\ref{lin}) is qualitatively
incorrect, since it predicts that the string is everywhere subsonic, that is,
$c_T<c_L,$ which we have seen is never the case. Models based on 
a self-dual equation of state predict that the string is everywhere
transonic, which is also false. 
These failures motivated attempts to construct improved actions 
\cite{CP,Ca2,HC},
of which the logarithmic approximation (\ref{log}) is an example.
As we have seen, the exact action for kinky vortons has a simple form
(\ref{lag}), but an action of this type does not appear to have
been suggested previously. Given the remarkable similarity between
the results obtained using (\ref{lag}) and numerical computations
in (3+1)-dimensions, it seems likely that an action of the form (\ref{lag})
will provide a good description for vortons.

\section{Intervals of instability}\news
For a circular elastic string of radius $R$, the frequencies $\Omega_n,$
of linear perturbations with fourier mode $n,$ can been related
to the transverse and longitudinal speeds \cite{CM}. Explicitly,
the scaled frequency $\nu_n=\Omega_n R/c_T$ satisfies the
cubic equation
\be
a_3\nu_n^3+a_2\nu_n^2+a_1\nu_n+a_0=0,
\label{cubic}
\ee
where
\bea
a_0&=&2(c_L^2-c_T^2)(n^2-1)n\\
a_1&=&4c_T^2(1-c_L^2)(n^2-1)-(1+c_T^2)(c_L^2-c_T^2)(n^2+1)\\
a_2&=&2c_T^2(c_L^2-c_T^2-2(1-c_L^2c_T^2))n\\
a_3&=&c_T^2(1+c_T^2)(1-c_L^2c_T^2).
\eea

The derivation of this formula in \cite{CM} assumes that the string moves
in a three-dimensional space, but the result applies equally well to a
string in two-dimensional space because the modes perpendicular to the
plane of the circular string decouple and are irrelevant for the 
stability analysis.

Instability is characterized by a complex root of the cubic (\ref{cubic}).
It is easy to show \cite{CM} that all roots are real for $n=0$ and $n=1,$ 
simply as a consequence of the causality restrictions $0<c_T^2\le 1$
and  $0<c_L^2\le 1$ for non-zero speeds.

For axially symmetric perturbations ($n=0$) the solutions of
the cubic are a trvial zero mode and
\be
\nu_0^2=\frac{2c_T^2(1-c_L^2)+(c_T^2+c_L^2)(1-c_T^2)}
{c_T^2(1+c_T^2)(1-c_L^2c_T^2)}.
\label{nu0}
\ee 
The explicit speed formulae (\ref{ct}) and (\ref{cl}) allow this
frequency to be calculated for any kinky vorton given the values of 
$\chi$ and $R.$ As an example, a kinky vorton is reported in \cite{BS}
with $Q=1500$ and $N=84,$ producing the quantities $\chi=0.0472$ and
$R=154.6.$ Using these values in the above formulae gives a frequency
$\Omega_0=0.00644.$ This can be compared with the results of full 
field dynamics, using the numerical approach described in detail in \cite{BS}.
Perturbing this kinky vorton by an axially symmetric mode that increases
the radius by $1\%$ produces an oscillation of the radius. 
A spectral analysis of this oscillation yields the  
frequency $0.00658,$ which is very close to the above predicted value
for $\Omega_0.$

For a given mode $n,$ the critical values of $\chi$ marking the limits
between stability and instability correspond to the values at which the
cubic (\ref{cubic}) has a repeated root. This is given by the vanishing
of the resultant
\be
\mbox{Res}(a_3\nu_n^3+a_2\nu_n^2+a_1\nu_n+a_0,3a_3\nu_n^2+2a_2\nu_n+a_1)
=27a_3^2a_0^2-18a_3^2a_0a_1a_2-a_3a_2^2a_1^2+4a_2^3a_3a_0+4a_1^3a_3^2.
\ee
 
\begin{figure}[ht]
\begin{center}
\includegraphics[width=12cm]{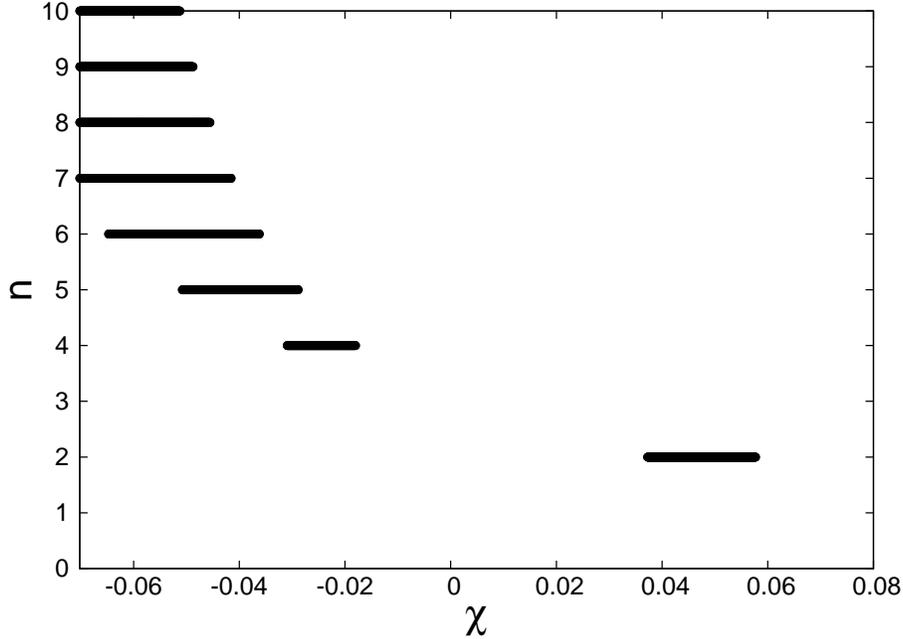}
\caption{
The $\chi$ intervals of instability for perturbations with fourier modes 
up to $n=10,$  for the theory with $\alpha=3/4.$
In this plot the lower limit of $\chi$ is equal to the value at which
the longitudinal speed vanishes.
}
\label{fig-chizone}
\end{center}
\end{figure}

In Figure~\ref{fig-chizone} we indicate the $\chi$ intervals of instability
for modes up to $n=10,$  for the theory with $\alpha=3/4.$ Note that 
the range of
$\chi$ for the plot runs from the value at which the longitudinal speed
vanishes $\chi_c=(3-\sqrt{17})/16$ (in which case there is automatically
instability). 

This plot reveals a complicated and highly non-trivial structure for
the instability intervals as a function of mode number. In particular,
if a kinky vorton is to be stable to all perturbation modes then there
is an extremely limited range in the magnetic regime ($\chi<0$). 

Note that the full field dynamical simulations of kinky vortons
presented in \cite{BS} included the perturbation of a solution with
$\chi=-0.04$ that did not decay. However, in this case the perturbation
was generated by the square boundary and therefore corresponds to modes
that are a multiple of $4.$ This result is consistent with
Figure~\ref{fig-chizone} since the unstable modes for $\chi=-0.04$ are
$n=5$ and $n=6.$ 

Most electric vortons ($\chi>0$) are stable to the instabilities 
discussed above and the results are consistent with full field
simulations \cite{BS}. The elastic string model predicts 
an instability for a small range of electric vortons around $\chi=0.05,$ 
under a perturbation with mode $n=2.$ A  
suitable kinky vorton solution to investigate this issue is the 
one discussed above with $Q=1500$ and
$N=84$ since this has $\chi=0.0472$ and $R=154.6.$ The elastic string model
predicts an instability to $n=2$ perturbations with the associated 
frequency $\Omega_2=0.0064+i0.00023.$ The period of kinky vorton 
oscillations is already large, but the imaginary part of 
$\Omega_2$ is a further order of magnitude smaller than the real part.
This implies that the timescale over which the instability
manifests itself is very large and long simulations are required to observe the
instability.

\begin{figure}[ht]
\begin{center}
\includegraphics[width=12cm]{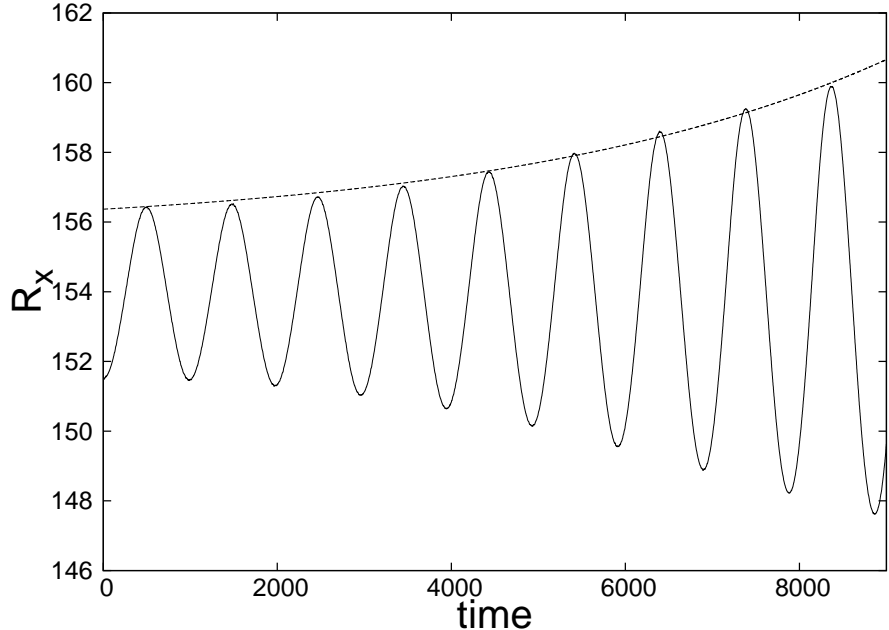}
\caption{
The evolution of the radius, $R_x$ (solid curve)
calculated along the $x$-axis, under an $n=2$ perturbation, for a kinky vorton
in a regime where instability is predicted.  
For comparison, the expected growth of the envelope is also
presented (dashed curve).
}
\label{fig-unstable}
\end{center}
\end{figure}

The numerical approach described in \cite{BS} has been applied
to study the stability of the above kinky vorton. The perturbation
consists of a squashing by $2\%$ along the $x$-axis and a stretching by the 
same factor along the $y$-axis, in order to preserve the total charge $Q.$
This elliptic deformation induces an $n=2$ perturbation and the
resulting evolution is presented in Figure~\ref{fig-unstable}, where
the solid curve represents the radius of the kinky vorton, as measured
along the $x$-axis. It is clear that the perturbation grows with time, 
confirming that the kinky vorton is unstable to modes with $n=2.$
The dashed curve displays the predicted growth rate of the envelope, 
that is, it is a curve with the predicted growth $\exp(t\Im(\Omega_2)).$
Once again, this shows an excellent agreement with the elastic string analysis.

The results described above, together with other 
 similar simulations, including
those presented previously in \cite{BS}, confirm the validity of the
elastic string description. The general conclusions are that most 
magnetic kinky vortons are unstable to generic non-axial perturbations, 
whereas the stability of electric kinky vortons has a crucial and 
complicated dependence upon the parameters of the kinky vorton.

\section{Conclusion}\news
In this paper we have derived an exact formula for the action density
on the string wordsheet in the kinky vorton model. Using this result
we have determined the equation of state and hence explicit 
expressions for the transverse and longitudinal propagation speeds in 
the elastic string description of a kinky vorton. This has allowed an
explicit analytic study of the stability of kinky vortons, revealing
a highly non-trivial pattern of intervals of instability. The analytic
results are in excellent agreement with numerical results from full
field simulations, confirming the validity of the elastic string
description.

The exact kinky vorton results share the same qualitative features
found in numerical computations of vortons, which suggests that the
kinky vorton form of the action density on the string worldsheet should
provide a good approximate description for vortons. 

A recent numerical investigation of vortons \cite{BS2} 
found instabilities to non-axial perturbations which are very 
similar to the instabilities described here for
kinky vortons. The kinky vorton results show that the existence of 
an instability is not a generic feature, but rather has a crucial
and non-trivial dependence on the properties of a particular kinky vorton.
Numerical simulations of vortons requires considerable computational
resources and unfortunately this restricts investigations to quite a 
limited region of parameter space. The results of the present paper
suggest that the instabilities found in vorton simulations \cite{BS2}
may not exist for all vortons. Certainly, an important observation is that
results found in a limited region of parameter space are unlikely to be 
generic throughout the parameter space of vorton solutions.

\section*{Acknowledgements}
PMS thanks the STFC for support under the rolling grant ST/G000433/1.

\end{document}